\pdfoutput=1
\documentclass[reprint,prb,aps,twocolumn,superscriptaddress,10pt]{revtex4-2}
\usepackage{amsmath}
\usepackage[english]{babel}
\usepackage{graphicx}
\usepackage{color}
\usepackage[dvipsnames]{xcolor}
\usepackage[normalem]{ulem}

\usepackage[colorlinks,
            citecolor = blue, 
						linkcolor = blue,
						urlcolor  = blue,
						anchorcolor = blue,
						breaklinks = true
						]{hyperref}

\usepackage{soul}
\usepackage{ amssymb }
\usepackage[exponent-product=\cdot]{siunitx}
\usepackage{physics}
\usepackage[dvipsnames]{xcolor}  

\newcommand{\mose}{MoSe$_2$}
\newcommand{\ws}{WS$_2$}
\newcommand{\hBN}{hBN}
\newcommand{\dop}{\ensuremath{\rho_\mathrm{AP}}}

\newcommand{\extfig}[1]{Extended Data Fig.~\ref{#1}}
\newcommand{\subfig}[2]{Fig.~\ref{#1}\textbf{#2}}
\newcommand{\subextfig}[2]{Extended Data Fig.~\ref{#1}\textbf{#2}}

\newif\ifdraft
\drafttrue

\def \ETHa{Institute for Quantum Electronics, ETH Z\"urich, CH-8093 Z\"urich, Switzerland}
\def \ETHb{Institute for Theoretical Physics, ETH Z\"urich, CH-8093 Z\"urich, Switzerland\\
$^\ddag$These authors contributed equally to this work}
\def \NIMSRCFM{Research Center for Electronic and Optical Materials, National Institute for Materials Science, 1-1 Namiki, Tsukuba 305-0044, Japan}
\def \NIMSICMN{Research Center for Materials Nanoarchitectonics, National Institute for Materials Science,  1-1 Namiki, Tsukuba 305-0044, Japan}

\def \BARCELONAa{Depeartament de F\'isica Qu\`antica i Astrof\'isica, Facultat de F\'isica, Universitat de Barcelona, E-08028 Barcelona, Spain}
\def \BARCELONAb{Institut de Ci\`encies del Cosmos, Universitat de Barcelona, ICCUB, Mart\'i i Franqu\`es 1, E-08028 Barcelona, Spain}

\def \TENNESSEEa{Department of Physics and Astronomy, University of Tennessee, Knoxville, TN 37996, USA}
\def \TENNESSEEb{Min H. Kao Department of Electrical Engineering and Computer Science, University of Tennessee, Knoxville, Tennessee 37996, USA}

\begin{document}


\title{Kinetic Magnetism in Triangular Moir\'e Materials}

\author{L.~Ciorciaro$^\ddag$}
\affiliation{\ETHa}

\author{T.~Smole\'nski$^\ddag$}
\affiliation{\ETHa}

\author{I.~Morera$^\ddag$}
\affiliation{\BARCELONAa}
\affiliation{\BARCELONAb}

\author{N.~Kiper}
\affiliation{\ETHa}

\author{S.~Hiestand}
\affiliation{\ETHa}

\author{M.~Kroner}
\affiliation{\ETHa}

\author{Y.~Zhang}
\affiliation{\TENNESSEEa}
\affiliation{\TENNESSEEb}


\author{K.~Watanabe}
\affiliation{\NIMSRCFM}

\author{T.~Taniguchi}
\affiliation{\NIMSICMN}

\author{E.~Demler}
\affiliation{\ETHb}

\author{A.~\.Imamo\u{g}lu}
\affiliation{\ETHa}

\maketitle



{\bf
Magnetic properties of materials ranging from conventional ferromagnetic metals to strongly
correlated materials such as cuprates originate from Coulomb exchange interactions. The existence
of alternate mechanisms for magnetism that could naturally facilitate
electrical control have been discussed
theoretically~\cite{Nagaoka1966,Haerter2005,Davydova2023,Morera2023,Lee2022,Carlstrom2022}
but an experimental demonstration~\cite{Dehollain2020} in an extended system has been missing.
Here, we investigate \mose/\ws\ van der Waals heterostructures in the vicinity
of Mott insulator states of electrons forming a frustrated triangular lattice and
observe direct evidence for magnetic correlations originating from a kinetic
mechanism. By directly measuring electronic magnetization through the strength of
the polarization-selective attractive polaron resonance~\cite{Sidler2016,Efimkin2017},
we find that when the Mott state is electron doped the system exhibits ferromagnetic
correlations in agreement with the Nagaoka mechanism.
}


Moir\'e heterostructures of two-dimensional materials provide a new
paradigm for the investigation of the physics of strongly correlated electrons.
In contrast to well-studied quantum materials, they provide a very high degree
of tunability of the parameters relevant for controlling correlations, such as
the carrier density and the ratio of interaction energy to hopping strength.
Moreover, unlike cold-atom quantum simulators, physics and functionality of
moir\'e materials can be varied using readily accessible external electric and
magnetic fields, creating a platform where different many-body phases compete.
During the five years since the first realization of a moir\'e material, a
wealth of correlation physics ranging from correlated Mott-Wigner states
through quantum anomalous Hall effect to superconductivity has been observed
both in magic angle twisted bilayer graphene (MATBG) and in bilayers of
transition metal dichalcogenides (TMDs)~\cite{Cao2018,Cao2018a,Yankowitz2019,Sharpe2019,Lu2019a,Serlin2020,
Wang2020,Shimazaki2020,Xu2020a,Regan2020,Li2021b,Li2021c,Huang2021,Tseng2022}.
With the exception of orbital magnetism
in MATBG~\cite{Lu2019a,Tschirhart2021,He2021} as well as early spin susceptibility
and scanning probe measurements in TMD bilayers~\cite{Tang2020a,Campbell2022,Tang2023,Foutty2023} however,
quantum magnetism in moir\'e materials has until recently remained experimentally
unexplored. Theoretical works on the other hand have investigated the magnetic
properties of the correlated Mott state in a moir\'e lattice with one electron
per lattice site~\cite{Hu2021,Morales-Duran2022} and focused on the possibility to realize quantum spin liquids~\cite{Balents2010,Morales-Duran2022,Kiese2022,Zhou2022}.

Here we investigate the magnetic properties of electrons in MoSe$_2$/WS$_2$
heterobilayers using low-temperature confocal microscopy. We focus on the
magnetization as a function of temperature $T$ and out-of-plane magnetic field $B_z$
at dopings around one electron per moir\'e lattice site ($\nu = 1$). For $\nu >
1$, our experiments show that the itinerant electrons exhibit a positive
Curie-Weiss constant $\theta_{CW}$. The abrupt jump in spin susceptibility as
soon as the electron density is tuned beyond $\nu=1$ at $T \approx \SI{170}{mK}$
is consistent with kinetic ferromagnetic correlations linked to the Nagaoka
mechanism~\cite{Nagaoka1966}.

\begin{figure*}[t]
    \includegraphics{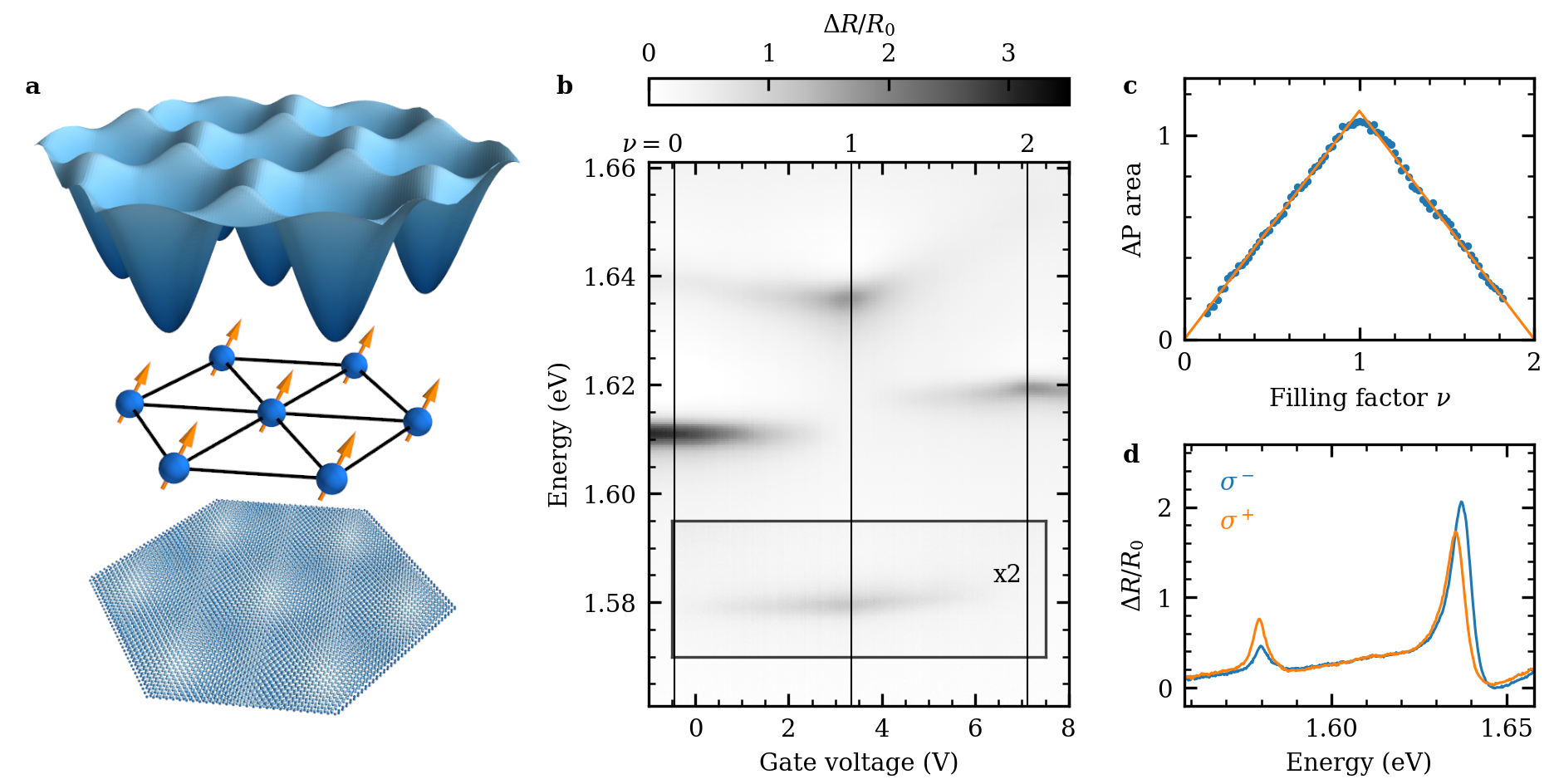}
    \caption{
    \textbf{a} Moir\'e potential in the conduction band. Electrons occupy the
    potential minima at the MM sites.
    \textbf{b} Normalized reflectance spectrum as function of gate
    voltage, tuning the doping of the heterostructure Sample II. At integer fillings,
    intensity maxima and cusps in the resonance energies appear.
    \textbf{c} Area of the AP resonance as function filling factor at $B_z = 0$.
    The linear increase and decrease confirm that electrons occupy a single
    minimum in the moir\'e unit cell, forming an isolated Hubbard band.
    \textbf{d} Polarization resolved reflection spectrum at $\nu = 1$, $B_z =
    \SI{1}{T}$, and $T = \SI{4.2}{K}$, Sample II. The AP resonance at \SI{1.58}{eV}
    is sensitive to the spin polarization of the electrons through its 
    degree of circular polarization.
    }
    \label{fig1}
\end{figure*}

We study two R-type \mose/\ws\ heterostructures encapsulated in \hBN. The
lattice mismatch and twist angle between the TMD monolayers creates a moir\'e
superlattice with a lattice constant of about \SI{7.5}{nm}. The minima of the
resulting electronic potential for the conduction band are located at the high
symmetry points where the metal atoms in the two layers are aligned (MM sites).
Injected electrons occupy the triangular lattice of MM sites as illustrated in
\subfig{fig1}{a}. In Sample I, the charge density and the electric field in the
heterostructure can be tuned independently using top and bottom graphene gates,
while Sample II is only single-gated.

The reflection spectrum as a function of electron density in \subfig{fig1}{b}
shows multiple resonances close to the energy of the optical transitions in
monolayer \mose. Intensity maxima and cusps in the resonance energies appear at
equally spaced gate voltages (see \extfig{extfig1} for extended range). These voltages
correspond to commensurate filling of the moir\'e superlattice with one or two
electrons per site ($\nu = 1$ and $2$, respectively), where incompressible
states are formed. We focus here on the resonance at
\SI{1.58}{eV}, which can be identified as an attractive polaron (AP) resonance
associated with collective excitation of bound electron-exciton pairs (trions)
located at the moir\'e potential minima~\cite{Suris2003,Sidler2016,Efimkin2017}.
As shown in \subfig{fig1}{c}, the area, or equivalently the oscillator strength, of
the AP resonance increases linearly as function of electron density up to $\nu =
1$ and subsequently decreases again linearly between $\nu = 1$ and 2. This
behaviour suggests the presence of an isolated Hubbard band where all electrons
occupy the same lattice sites, forming double occupancies (doublons) for $\nu > 1$. Since the AP resonance is
associated with the bound trion of an exciton and a resident electron, it can
only be optically excited on moir\'e lattice sites already occupied by a single
electron. Consequently, the densities $\nu = \varepsilon$ and $\nu = 2 - \varepsilon$ provide
the same number of sites for AP formation and hence lead to an identical
oscillator strength of the AP resonance.

The optical selection rules of monolayer \mose\ are retained in the
heterostructure, giving rise to circularly polarized resonances for $B_z \neq 0$,
corresponding to transitions in the K and K' points of the \mose\ Brillouin zone~\cite{Yao2008}. 
The linear dependence of the AP peak area on the electron density, together with
the optical valley selection rules and strong spin-orbit coupling leading to
spin-valley locking, allows us to employ the polarization resolved AP resonance as
a quantitative probe of the degree of spin polarization of the electrons.  Since
the AP is only formed by excitons in the K valley and spin-down electrons in the
K' valley or vice versa~\cite{Suris2003}, the AP oscillator strength in
$\sigma^+$-polarization ($\sigma^-$-polarization) is proportional to the density
$n_\downarrow$ ($n_\uparrow$) of spin-down (spin-up) electrons.  The degree of
spin polarization is then given by
\begin{equation}
    \rho_s = \frac{n_\uparrow - n_\downarrow}{n_\uparrow + n_\downarrow}
    = \frac{A_{\sigma^-} - A_{\sigma^+}}{A_{\sigma^-} + A_{\sigma^+}}
    =: \dop,
\end{equation}
where $A_{\sigma^\pm}$ is the area of the AP resonance in $\sigma^\pm$ polarization
and \dop\ denotes the degree of circular polarization of the AP resonance.
The polarization resolved spectrum in \subfig{fig1}{d}, measured at $B_z =
\SI{1}{T}$, $T = \SI{4.2}{K}$, and $\nu = 1$, highlights how the AP resonance
becomes partially polarized in a moderate magnetic field. Note that the
resonance at \SI{1.635}{eV} is also sensitive to the spin polarization, mainly
through a splitting with giant effective g-factor $g_\mathrm{eff} = 31$, as
previously reported for other moir\'e heterostructures~\cite{Tang2020a,Campbell2022}.

\begin{figure}
    \includegraphics{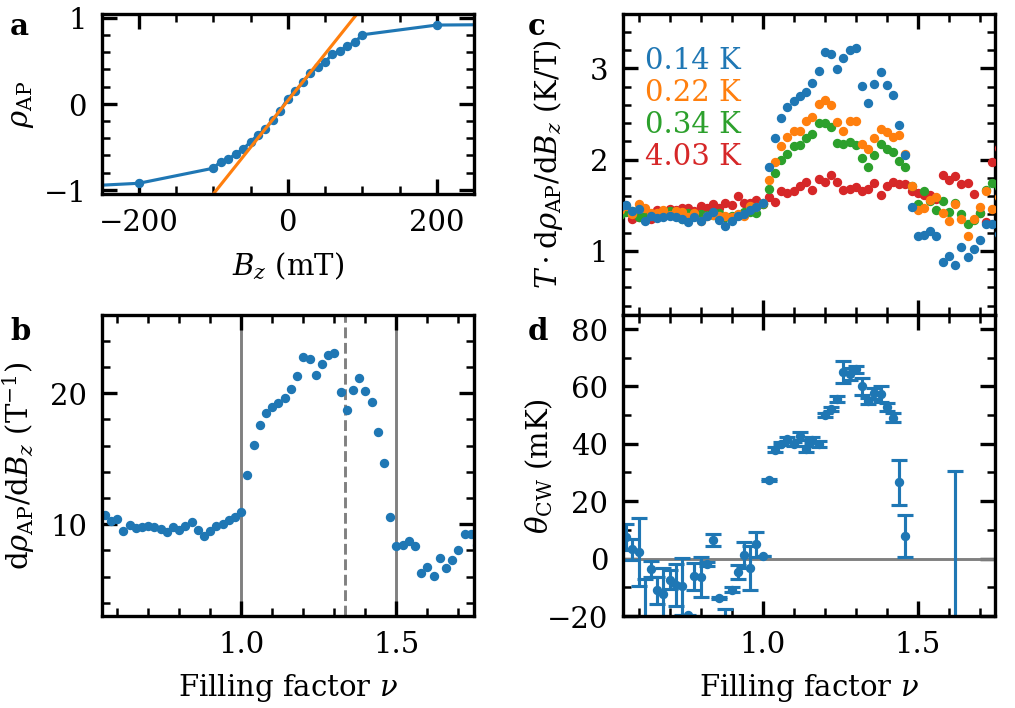}
    \caption{ Low temperature data, Sample I.
    \textbf{a} Degree of polarization of the AP resonance as function of
    magentic field and linear fit around $B_z = 0$ yielding the susceptibility.
    \textbf{b} Slope of \dop\ as function of electron filling
    factor. A sharp increase in spin susceptibility is observed at $\nu = 1$.
    \textbf{c} Doping dependence of $\dd\dop/\dd B$ at different temperatures,
    normalized by $1/T$. At $1 < \nu < 1.5$, the susceptibility deviates from
    $1/T$ due to ferromagnetic interactions.
    \textbf{d} Fitted Curie-Weiss constant as function doping. 
    Error bars correspond to the standard error of the fit.
    }
    \label{fig2}
\end{figure}

In order to gain insight into the interactions between spins of the electrons
residing in the superlattice potential, we measure the AP degree of polarization
\dop\ as a function of $B_z$ for
electron densities satisfying $0.5 < \nu < 1.8$. An excitation power of \SI{4.7}{pW}
is used to avoid laser-induced spin depolarization (see Methods) and thereby ensure
that we probe magnetic properties of the electronic ground state~\cite{Wang2022a}. We perform a linear fit
to extract the slope at $B_z = 0$, as shown in \subfig{fig2}{a}, which is related to the
magnetic susceptibility through
\begin{equation}
\frac{\mathrm{d}}{\mathrm{d}B_z}\dop(\nu) 
= \frac{\mathrm{d}}{\mathrm{d}B_z} \frac{M(\nu)}{M_s(\nu)} 
= \frac{\chi(\nu)}{\mu_0 M_s(\nu)},
\end{equation}
where $M(\nu)$ is the magnetization, $\mu_0$ the vacuum permeability, and $M_s(\nu) =
g \mu_\mathrm{B}n_{e,\nu}\left(1 - \left| \nu - 1 \right|\right) / 2$ the saturation
magnetization for each density, with $\mu_\mathrm{B}$ the Bohr magneton, $g$ the
\mose\ conduction band g-factor, and $n_{e,\nu}$ the electron density at $\nu=1$.
The slope $\mathrm{d}\dop/\mathrm{d}B_z$ measured at $T \approx \SI{170}{mK}$ as
function of the filling factor $\nu$ is shown in \subfig{fig2}{b}. It is
constant at $\nu < 1$ and has a sharp increase at $\nu = 1$, where the system
transitions from a hole-doped to an electron-doped Mott insulator.
Strikingly, there is a similarly sharp decrease again at $\nu = 1.5$, even
though this is not a characteristic filling of the triangular lattice and an ordered
state underlying the abrupt change can only emerge with additional symmetry
breaking. We focus here on the density dependence of magnetic interactions 
around $\nu = 1$ and  further investigate them by performing temperature-dependent
measurements shown in \subfig{fig2}{c}. Each curve measured at a temperature $T$ is 
normalized by $1/T$ such that for paramagnetic behaviour
the curves collapse onto one value, as observed for $\nu < 1$. The
deviation from this universal behaviour for $\nu > 1$ is evidence for the
presence of ferromagnetic interactions.

In general, exchange interactions are expected to play a key role in determining the magnetic order of
the system. For a moir\'e structure with lattice constant of \SI{7.5}{nm} in particular,
first principle calculations predict reduced superexchange due to the large on-site
repulsion, such that direct exchange is the relevant interaction to consider~\cite{Morales-Duran2022}.
The magnetic properties at $\nu=1$, where 
the electrons form an incompressible Mott insulator and are localized on moir\'e lattice sites, should be exclusively determined by exchange interactions.
Surprisingly, we do not find a significant deviation from paramagnetic behaviour
at $\nu = 1$, as can be seen in \subfig{fig2}{d} from the fitted Curie-Weiss
constant $\theta_\mathrm{CW}$ as function of doping. This suggests that exchange
interactions do not play a significant role in our system.
Possible explanations for this  are a cancellation of competing direct and superexchange
interactions or that the electrons are more strongly localized than predicted by theory due
to a deeper moir\'e potential. 

In addition to exchange, effective magnetic interactions based on a kinetic
mechanism appear when the electrons are mobile~\cite{Xu2022}: minimization of the kinetic energy gives rise
to magnetic order in a Hubbard band close to half filling ($\nu=1$) even in the limit of
vanishing spin interactions~\cite{Nagaoka1966}.  We attribute the observed ferromagnetic
correlations with a sharp onset at $\nu = 1$ to kinetic magnetism. This is corroborated by
the dip in susceptibility at $\nu = 4/3$, marked by the dashed line in \subfig{fig2}{b}:
At fractional filling factors commensurate with the moir\'e superlattice,
electrons have been shown to form incompressible Mott-Wigner states
\cite{Regan2020,Xu2020a,Huang2021,Li2021c,Tang2023}, reducing their mobility and
suppressing kinetic magnetism.



\begin{figure}
    \includegraphics{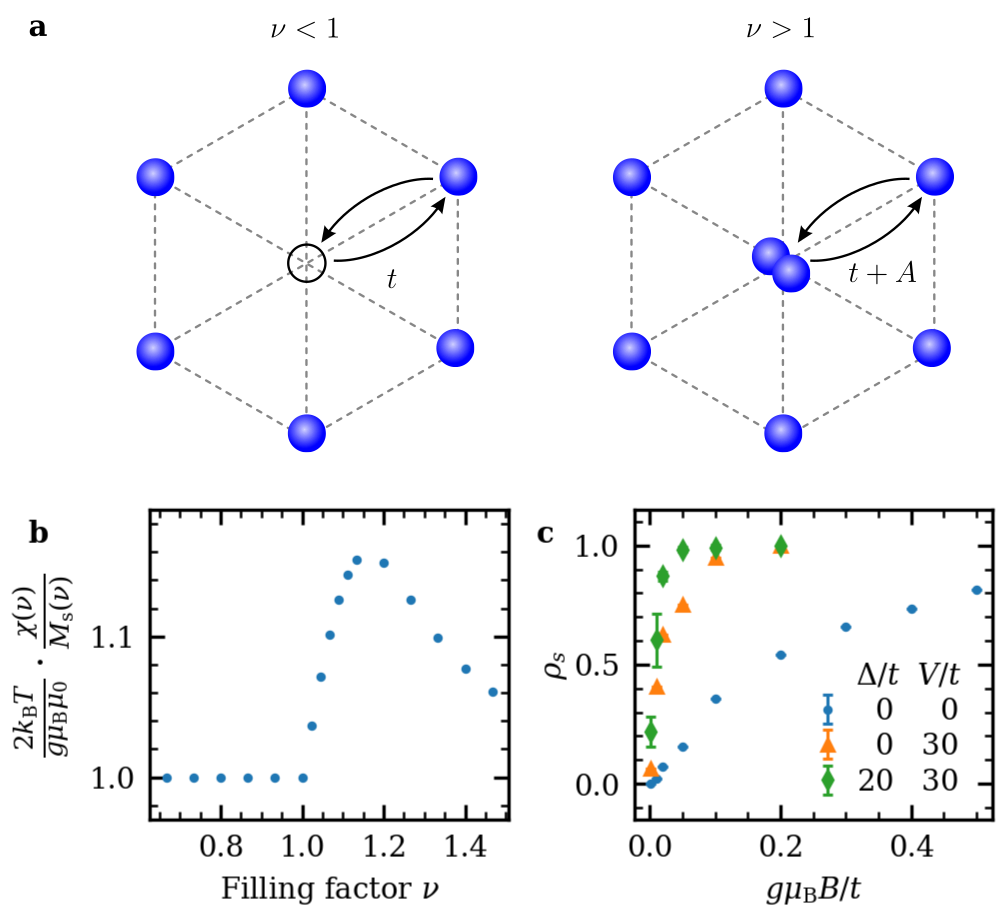}
    \caption{
    \textbf{a} Illustration of hopping processes for holes and doublons. 
    The presence of long-range interactions introduces an assisted hopping
    $A$ that modifies the doublon hopping strength.
    \textbf{b} Simulated spin susceptibility as function of filling factor for
    assisted hopping $A = 10t$ and temperature $k_\mathrm{B}T = A$, but absent
    long-range Coulomb interaction $V=0$, disorder $\Delta=0$, and exchange
    interactions $J=0$. The assisted hopping leads to an asymmetry between
    doublons and holes. The normalization is chosen such that the value $1$
    corresponds to a paramagnetic response.
    \textbf{c} Simulated degree of spin polarization as function of magnetic field at $\nu
    = 0.89$ and $T = 0$. Both disorder with distribution width $\Delta$ and
    long-range Coulomb interactions $V$ suppress the antiferromagnetic
    correlations.
    }
    \label{fig3}
\end{figure}

In a simple single-band Fermi-Hubbard model in the strongly interacting regime, the
kinetic mechanism leads to a transition from antiferromagnetic to ferromagnetic
interactions close to $\nu=1$ on a triangular
lattice~\cite{Morera2023,Lee2022,Carlstrom2022}. To qualitatively
understand the experimentally observed absence of antiferromagnetic correlations for $\nu < 1$,
we consider an extended model (see Methods) taking into account long-range Coulomb interactions
$\hat{V}$, while setting exchange interactions $J$ to zero, motivated by the  paramagnetic
response at $\nu=1$.
The Coulomb interaction term
modifies the hopping of electrons onto sites that are already occupied, which
renormalizes the doublon hopping, while leaving the hole hopping $t$ unchanged,
as illustrated in \subfig{fig3}{a}. The effective doublon hopping is given by
$t + A$, with the assisted hopping term
$A = - \left<w_i, w_i\right| \hat{V} \left|w_i, w_j\right>$,
where $w_i$ and $w_j$ denote states localized on neighbouring sites~\cite{Hu2021,Morales-Duran2022}.
Due to the asymmetry in hopping between holes and doublons, the kinetic
magnetism is enhanced for $\nu > 1$.
Therefore, in an intermediate temperature range 
$t\ll k_BT\lesssim t + A$, we expect a sizeable modification of the susceptibility for
$\nu > 1$, but only negligible deviations from a paramagnetic response for $\nu \leq 1$.
This asymmetry of the susceptibility around $\nu = 1$ is captured by our finite
temperature tensor network simulations shown in \subfig{fig3}{b}. Details on the theoretical
model employed and the simulations can be found in the Methods section.

Furthermore, we find that the overall hopping strength is renormalized by the presence of long-range interactions and$/$or disorder, reducing the strength of kinetic magnetism, particularly for $\nu < 1$.
In \subfig{fig3}{c} we show simulated magnetization curves at $\nu = 0.89$ and $T = 0$
comparing the cases with and without long-range interactions or disorder. Introducing
disorder or interactions leads to an increased slope at low fields, corresponding to
an enhanced susceptibility or suppressed antiferromagnetic correlations.
This limits the temperatures required to observe kinetic magnetism in the moir\'e structure
to smaller values than what would be expected from theoretically predicted hopping strengths
on the order \SI{1}{meV} for such moir\'e structures.

\textit{Conclusion}---We have demonstrated accurate determination of the spin polarization through polarized AP oscillator
strength measurements at picowatt power levels. We have
used this method to study the density-dependent spin susceptibility and found a sudden
appearance of ferromagnetic correlations for $\nu > 1$. Supported by tensor network
simulations, our observations can be attributed to kinetic magnetism in an
extended single-band Fermi-Hubbard model on a triangular lattice. 
Even though prior studies found good agreement with ab initio calculations,
our observation of a paramgentic response at $\nu=1$ suggests that direct exchange and superexchange
interactions are weak and the spin state is dominated by effective kinetic interactions.  The strong
asymmetry between $\nu<1$ and $\nu>1$ suggests the presence of a large Coulomb-assisted hopping of
doublons. Moreover, Coulomb interaction and disorder renormalize the effective hopping for
holes and doublons, which in turn reduces the scale of magnetic correlations to low
temperatures below \SI{100}{mK}.

\textit{Note added}---During the preparation of this manu\-script, we became aware
of several parallel works exploring different aspects of magnetism in bilayer MoTe$_2$ moir\'e
structures~\cite{Cai2023,Anderson2023,Zeng2023}.



\clearpage

\renewcommand{\figurename}{Extended Data Figure}
\setcounter{figure}{0}

\onecolumngrid

\begin{figure*}
    \includegraphics{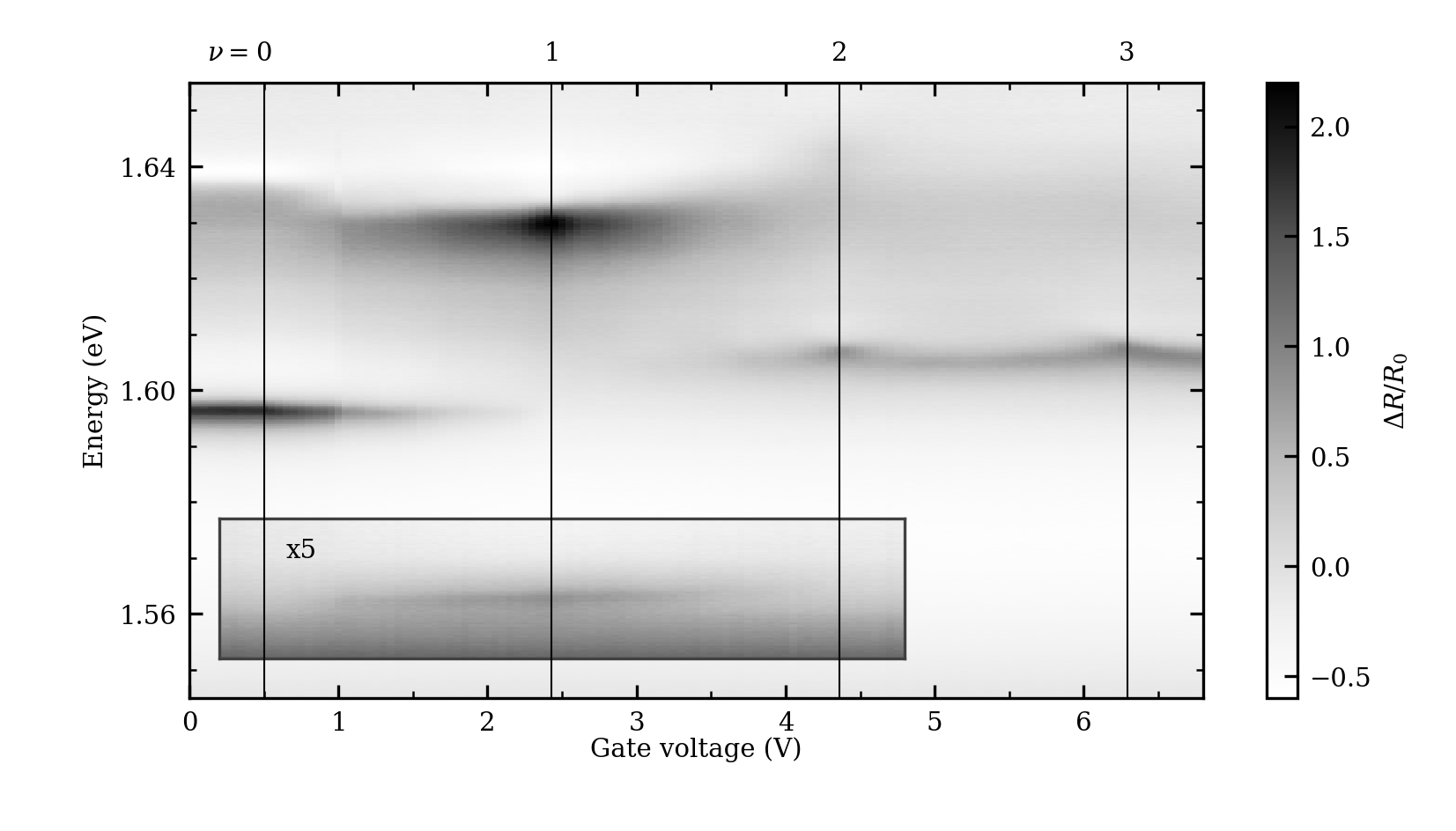}
    \caption{Normalized reflectance spectrum as function of doping, Sample I, extended
    range.}
    \label{extfig1}
\end{figure*}

\twocolumngrid

\section*{Methods}
\subsection*{Sample Fabrication}
Graphene, hBN, \mose, and \ws\ were exfoliated from bulk crystals onto
Si/SiO$_2$ (285\,nm) substrates and assembled in heterostructures using a
standard dry-transfer technique with a poly(bisphenol A carbonate) film on a
polydimethylsiloxane (PDMS) stamp. Both samples were encapsulated between two
approximately \SI{30}{nm} thick hBN flakes. Optical lithography and electron
beam metal deposition were used to fabricate electrodes for the electrical
contacts. 

\begin{figure}
    \includegraphics{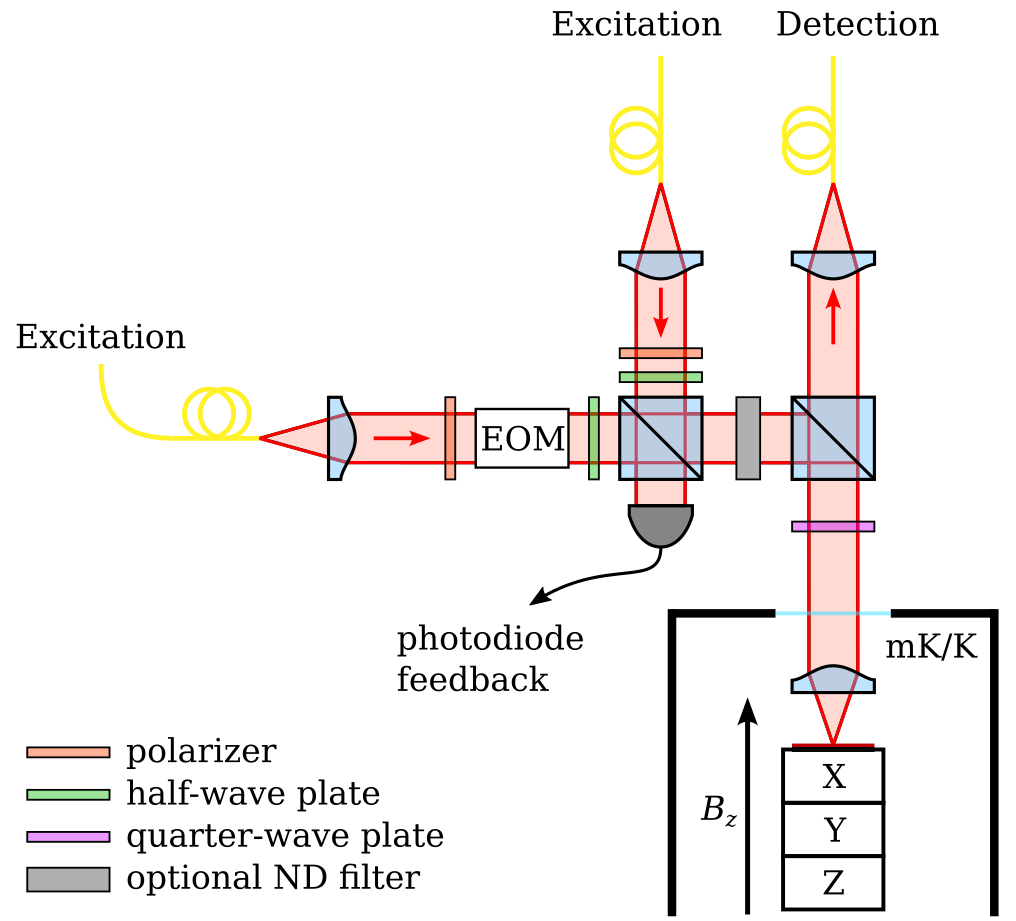}
    \caption{Schematic of the optical setup.
    }
    \label{extfig2}
\end{figure}

\subsection*{Experiment Setup}
Experiments were done in a liquid helium bath cryostat and a dilution
refrigerator with free-space optical access, with windows at the still,
\SI{4}{K}, and room temperature stages. The samples
were mounted on three piezoelectric nanopositioners. A fiber-coupled confocal
microscope was used for optical measurements, with a either a single aspheric
lens or an objective ($\mathrm{NA} = 0.7$ for both) focusing the light to a
diffraction-limited spot. A schematic of the optical setup is shown in
\extfig{extfig2}. Reflection spectra were measured using a supercontinuum laser
with a variable filter as light source and a spectrometer with a liquid-nitrogen-
or Peltier-cooled CCD camera as detector. For resonant single-frequency
measurements of the magnetic circular dichroism (MCD), we used a tunable CW
titanium sapphire laser and a Geiger-mode avalanche photodiode for detection of
low-power signals. The laser polarization was switched between $\sigma^+$ and
$\sigma^-$ at kilohertz rates using an electro-optico modulator in order to
reduce the sensitivity of the MCD measurement to slow drifts. All measurements
were power-stabilized with feedback from a photodiode to an acousto-optic
modulator or a fiber-coupled variable optical attenuator.

\begin{figure*}[t]
    \includegraphics{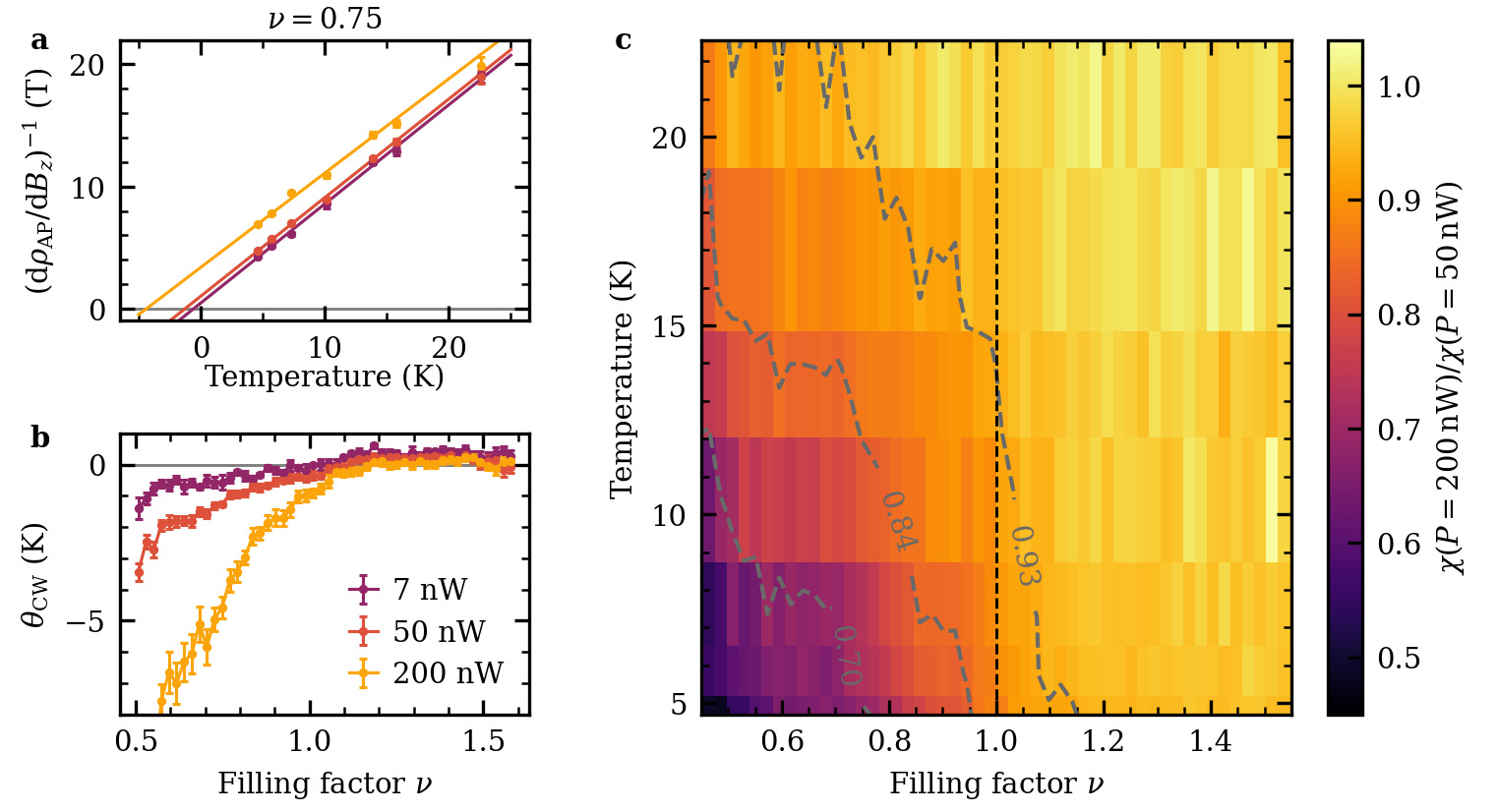}
    \caption{High temperature data, Sample II.
    \textbf{a} Inverse slope $\dd\dop/\dd B$ measured at $\nu=0.75$ as function of
    temperature for different excitation powers. For each power, the data are precisely reproduced by the
    Curie-Weiss formula with the Curie-Weiss constant getting sizably lower for larger powers due to
    light-induced electron spin depolarization.
    \textbf{b} Fitted Curie-Weiss constant as function of electron filling
    factor for three different excitation powers. For $\nu < 1$, the light-induced
    spin depolarization leads to apparent negative Curie-Weiss constants.
    \textbf{c} Ratio of the slope $\dd\dop/\dd B_z$ measured at two
    different powers, plotted as function of temperature and filling factor.
    The power dependence changes with both temperature and filling factor and is
    most prominent at small $\nu$ and low temperatures. The grey dashed lines mark contours of fixed
    susceptibility ratios (as indicated). Error bars correspond to the standard error of the fit.
    }
    \label{extfig3}
\end{figure*}

\subsection*{Background Subtraction}
Differential reflectance presented in the plots is defined as $\Delta R / R_0 =
(R - R_0) / R_0$, where $R$ is the measured reflection spectrum of the
heterostructure and $R_0$ is the background reflection spectrum on the hBN
flakes away from the TMD flakes. For the MCD measurements, the background
reflectance at the laser frequency is measured in both polarizations at charge
neutrality or high electron density ($\nu > 2$) where there is no AP resonance.
The degree of circular polarization is then given by
\begin{equation}
    \dop = \frac{( R^{\sigma^+} - R_0^{\sigma^+} ) - 
                 ( R^{\sigma^-} - R_0^{\sigma^-} )}
                {( R^{\sigma^+} - R_0^{\sigma^+} ) + 
                 ( R^{\sigma^-} - R_0^{\sigma^-} )}.
\end{equation}

\subsection*{Power Dependence of Spin Polarization}
To access the true magnetic ground state properties of the system, it is
essential to ensure that the intensity of the probe light is sufficiently low to
not perturb the system. Similarly to monolayer MoSe$_2$~\cite{Smolenski2022},
light illumination leads to depolarization of the spin population in the moir\'e
heterostructure. Since the strength of the depolarizing effect depends on both
temperature and charge density, it can give rise to misleading artefacts in the
measured electronic magnetism. This is directly revealed by our
filling-factor-dependent measurements of the Curie-Weiss constant carried out
at high temperatures $T>\SI{4}{K}$ in the bath cryostat on Sample II. In these
experiments, the sample was illuminated with a white light of tunable power. The
magnetic susceptibility of the electron system was extracted based on the
degree of circular polarization of the AP resonance that was in
turn determined by fitting its spectral profile with a dispersive Lorentzian
lineshape~\cite{Smolenski2021}. On this basis, we were able to analyze the
temperature dependence of the inverse magnetic susceptibility for various
filling factors and excitation powers. As seen in \subextfig{extfig3}{a} (for
$\nu=0.75$), even though the employed powers remain in the sub-microwatt range, they
still markedly affect the magnetic response. More specifically, the Curie-Weiss
constant is clearly lower for larger excitation powers. This effect is most
prominent for low filling factors and becomes indiscernible at $\nu\gtrsim1$
(see \subextfig{extfig3}{b}). 

This power dependence originates primarily from the changes in spin-valley
relaxation dynamics of the electron system. As it has been demonstrated in prior
studies of TMD monolayers~\cite{Li2021d}, the spin relaxation time becomes
shorter for larger electron densities and higher temperatures. As a result, if a
certain number of electrons undergo a spin flip due to the interaction with
optically injected excitons, it takes longer for them to relax back to their
ground state when $\nu$ and $T$ are low. For this reason, the magnetic
susceptibility determined upon exciton injection into the system is lower compared
to its unperturbed value. Moreover, the deviation between these two quantities
becomes larger for higher excitation powers and lower $\nu$ and $T$ (see
\subextfig{extfig3}{c}), which explains the striking power dependence of the
Curie-Weiss constant at $\nu<1$ in \subextfig{extfig3}{b}. In particular,
the data in this figure directly reveal that in order for the excitons to
constitute a nondestructive probe of the electron spin system at $T>\SI{4}{K}$ and
$0.5\lesssim\nu\lesssim1.5$, the excitation power needs to be around a few nW. 

\begin{figure}
    \includegraphics{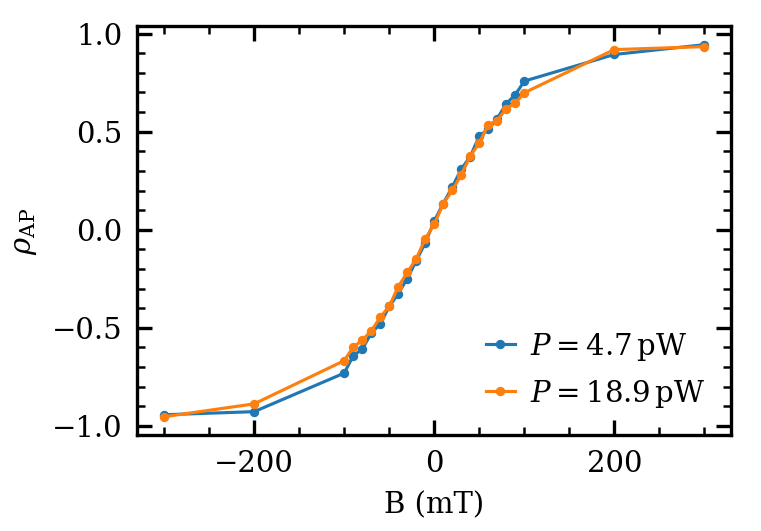}
    \caption{Magnetization curves measured with two excitation powers at
    $T \approx \SI{170}{mK}$ in Sample I. The curves measured at powers differing
    by a factor of 4 overlap, indicating that the exciation power is sufficiently
    low to not perturb the system.}
    \label{extfig4}
\end{figure}

Owing to the aforementioned temperature dependence of the spin-valley relaxation
time, accessing the true magnetic ground state properties of the electron system
at mK temperatures requires us to further reduce the excitation power.
We find that the requisite power is on the order of a few pW, as seen in
\extfig{extfig4}, where increasing the light intensity by a factor of
four did not affect the outcome of the measurement and the two
magnetization curves overlap.
Taking this into account, we used a resonant laser with \SI{4.7}{pW} incident power
on the sample in our mK measurements. Note that this level of power is about six orders
of magnitude below the level at which the laser measurably heats the cold finger
in the cryostat. Because the line shape and energy of the AP are almost constant
with respect to gate voltage and magnetic field, measuring the reflectance at a
single frequency is equivalent to measuring the area of the peak.

\subsection*{Temperature Calibration}

\begin{figure}
    \includegraphics{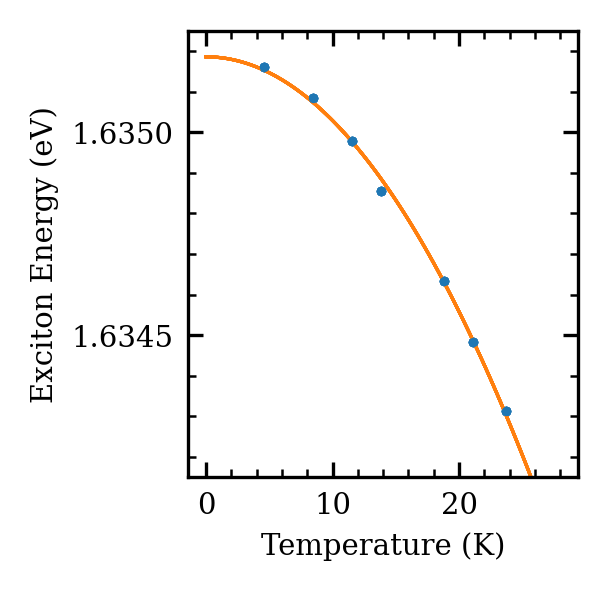}
    \caption{\mose\ exciton energy as function of temperature. The energy
    follows the quadratic behaviour $E(T) = E_0 - \gamma T^2$ with fitted
    parameters $E_0 = \SI{1.6352}{eV}$ and $\gamma = \SI{1.57}{\micro eV/K^2}$.}
    \label{extfig5}
\end{figure}

To calibrate the lowest temperatures used in the dilution refrigerator, we first
carried out Curie-Weiss fits for $T > \SI{300}{mK}$ relying on the built-in
temperature readout based on a resistance measurement. From these fits,
we find Curie-Weiss constants close to 0 for $\nu \le 1$, which allows us to
assume paramagnetic behaviour $\dd\dop(T)/\dd B_z = g\mu_\mathrm{B} / (2
k_\mathrm{B} T)$, where $\mu_\mathrm{B}$ is the Bohr magneton, $k_\mathrm{B}$
the Boltzmann constant, and $g$ the electronic g-factor. This is further
confirmed by measured magnetization curves that follow $\dop(B_z) =
\tanh(g\mu_\mathrm{B}B_z/(2k_\mathrm{B}T))$. The value $g = 4.5$ of the g-factor
can be fixed from this relation using the measured magnetization slope and
temperature.  We then use the same relation to extract the temperature at $T <
\SI{300}{mK}$ using the measured slope at $\nu = 1$.


The slope of $\dd\dop(T)/\dd B_z$ at $\nu=1$ was also utilized to determine the
sample temperature in high-temperature measurements that were performed in the
bath cryostat. In this case, the obtained temperature values were further
verified by analysis of the temperature-induced redshift of the exciton
resonance in a MoSe$_2$ monolayer region of Sample II. As shown in
\extfig{extfig5}, the measured energy $E_X(T)$ of this resonance
decreases quadratically with temperature, following aptly the Varshni formula
$E_X(T)=E_0-\gamma T^2$~\cite{Varshni1967}. The corresponding
$\gamma=\SI{1.6}{\micro eV/K^2}$ agrees very well with the values reported in
previous studies of \mose\ monolayers carried out in wider
temperature ranges~\cite{Arora2015}. This finding provides a strong confirmation
of the validity of our temperature calibration procedure.

\subsection*{Theoretical Model}
To explain the experimental results we consider a single-band extended $t$-$J$ model,
\begin{align}
    \hat{H} = &-t\, \hat{P}\sum_{\langle i,j\rangle,\sigma } \left(\hat{c}_{i,\sigma}^{\dagger}\hat{c}_{j,\sigma} + \textrm{h.c.}  \right)\hat{P} \\
    &+ J \sum_{\langle i,j\rangle}\left(\mathbf{S}_i\mathbf{S}_j -\frac{1}{4}\hat{n}_{i}\hat{n}_{j}\right) \nonumber\\
    &-\frac{A}{2}\, \hat{P}\sum_{\langle i,j\rangle,\sigma } \left[\hat{c}_{i,\sigma}^{\dagger} \left(\hat{n}_{i,\bar{\sigma}} + \hat{n}_{j,\bar{\sigma}}\right) \hat{c}_{j,\sigma}  + \textrm{h.c.} \right]\hat{P} \label{Eq:tJ} \\
    &+ V\sum_{i<j}\frac{\hat{n}_{i}\hat{n}_{j}}{|i-j|}-h\sum_{i}\hat{S}^z_i+\sum_i \Delta_i \hat{n}_i, \nonumber
\end{align}
where $t$ is the hopping strength, $J$ is the spin-spin interaction, $A$ is the assisted hopping, 
$V$ is the strength of Coulomb interaction projected into the lowest Wannier orbital, $h$ is
the external magnetic field in units of $g\mu_B$, $\Delta_i$ is the on-site potential
energy, and $\hat{P}$ is a projector that projects out doublons (holes) in the hole (electron)
doped regime. We consider a null spin-spin interaction $J=0$ motivated by the experimental
results pointing to a paramagnetic response at $\nu=1$. Moreover, to
implement the long-range coupling proportional to $V$ we cut the range of the interaction
at third neighbours. The on-site potential energy $\Delta_i$ takes into account the
spatial variations of the moir\'e potential. We consider a uniformly distributed disorder
$\Delta_i \in [-\Delta/2, \Delta/2 )$ of width $\Delta$ with a corresponding RMS parameter $\Delta/\sqrt{12}$. 

\subsection*{Model Parameters}
To estimate the relevant parameters of the Hamiltonian used in the tensor network
simulations, we start from the finite discrete Fourier expansion of the moir\'e potential,
\begin{equation}
V(\vec{r}) =  \sum_{n=1}^{6} V_n e^{i \vec{G}_n \cdot \vec{r}},
\end{equation}
where $V_n=-V_0 \exp \left[i(-1)^{n-1} \varphi\right]$ and we introduce the reciprocal lattice vectors
\begin{equation}
\vec{G}_n = \frac{4\pi}{a_\mathrm{m}\sqrt{3}}
\begin{pmatrix}
    \cos(\pi n / 3) \\
    \sin(\pi n / 3)
\end{pmatrix}.
\end{equation}
The parameters $V_0 = \SI{6.3}{meV}$ and $\varphi = 0$ are obtained from first
principles calculations.

The single-electron problem is described by the low-energy Hamiltonian,
\begin{equation}
    \hat{H} = -\frac{\hbar^2}{2m^*}\vec{\nabla}^2 + \hat{V}(\vec{r}),
\end{equation}
where we introduce the effective mass $m^*=0.7 m_e$ of the \mose\ conduction band electrons.
Since the moir\'e potential has a periodic structure we can employ Bloch's theorem to write the wavefunctions as
\begin{equation}
    \psi^{(n)}_{\vec{k}}(\vec{r}) = u^{(n)}_{\vec{k}}(\vec{r})e^{i \vec{k}\cdot \vec{r}},
\end{equation}
where $n$ is the band index, $\vec{k}$ is restricted to the first moir\'e Brillouin zone, and
$u^{(n)}_{\vec{k}}$ are the Bloch functions. Since the Bloch functions have the same
periodicity as the moir\'e potential, 
$u^{(n)}_{\vec{k}}(\vec{r}) = u^{(n)}_{\vec{k}}(\vec{r}+\vec{R}_i)$,
we can expand them by performing a discrete Fourier transform
\begin{equation}
    u^{(n)}_{\vec{k}}(\vec{r}) = \sum_{\vec{G} \in \mathcal{G}} c^{(n)}_{\vec{k},\vec{G}}e^{i\vec{G}\cdot \vec{r}},
\end{equation}
where $\mathcal{G}$ is the set of all reciprocal lattice vectors. Therefore,
the Hamiltonian can be written in the basis of reciprocal lattice vectors as
\begin{equation}
    H_{\vec{G},\vec{G}'}(\vec{k}) = \frac{\hbar^2}{2m^*}\left(\vec{k}+\vec{G} \right)^2 \delta_{\vec{G},\vec{G}'} +  \sum_{n=1}^{6}V_n \delta_{\vec{G}-\vec{G}',\vec{G}_n},
\end{equation}
which can be diagonalized for each quasi-momentum $\vec{k}$ by using a large set of reciprocal lattice vectors. The ground state solution corresponds to the lowest band $n=0$. The associated Wannier wavefunction $w_i(\Vec{r})$ localized at site $\vec{R_i}$ is obtained by performing the change of basis
\begin{equation}
    w_i(\Vec{r}) = \frac{1}{\sqrt{\mathcal{N}}} \sum_{\vec{k}\in\mathrm{BZ}} \psi_{\Vec{k}}(\Vec{r}) e^{i \Vec{k}\cdot \Vec{R}_i},
\end{equation}
where we drop the band index and introduce the normalization factor $\mathcal{N}$. 
    
The interaction potential between charges within the TMDs is given by the Rytova-Keldysh potential~\cite{Rytova1967,Keldysh1979}
\begin{equation}
V_\mathrm{RK}(r) = \frac{e^2}{8\epsilon_0 r_0} \left(
H_0\left(\frac{\epsilon_\mathrm{r} r}{r_0}\right) -
Y_0\left(\frac{\epsilon_\mathrm{r} r}{r_0}\right)\right),
\end{equation}
where $H_0$ is the Struve function, $Y_0$ the Bessel function of the second
kind, $r_0 = \SI{3.5}{nm}$ the screening length for \mose, and
$\epsilon_\mathrm{r} = 4.5$ the relative permittivity of hBN as the surrounding
medium~\cite{Goryca2019}. The matrix elements
\begin{align}
t & = -\bra{w_i} \hat{H} \ket{w_j} = \SI{0.75}{meV} \nonumber\\
U & = \bra{w_i, w_i} V_\mathrm{RK}(|\vec{r_2} - \vec{r_1}|) \ket{w_i, w_i} = \SI{157}{meV} \nonumber\\
V & = \bra{w_i, w_j} V_\mathrm{RK}(|\vec{r_2} - \vec{r_1}|) \ket{w_i, w_j} = \SI{44.6}{meV}\\
J & = -\bra{w_i, w_j} V_\mathrm{RK}(|\vec{r_2} - \vec{r_1}|) \ket{w_j, w_i} = \SI{-0.61}{meV} \nonumber\\
A & = -\bra{w_i, w_i} V_\mathrm{RK}(|\vec{r_2} - \vec{r_1}|) \ket{w_i, w_j} = \SI{6.1}{meV} \nonumber
\end{align}
are evaluated numerically, where $\ket{w_i,w_j}$ denotes a state where two
electrons occupy neighbouring Wannier orbitals.

\subsection*{Tensor Network Simulations}
Our finite temperature tensor network simulations are based on a purification scheme performed in the canonical ensemble. We implement the $U(1)$ symmetry associated with the conservation of the total number of electrons, but we do not fix the net magnetization of the system. The finite temperature density matrix is represented as a matrix product state (MPS) in a doubled Hilbert space. The MPS maximum bond dimension is set to $\chi=768$. The cooling process is performed similarly to Ref.~\cite{Morera2023}. We progressively apply an infinitesimal ($\delta \beta=0.1$) Boltzmann factor $e^{-\delta \beta/2}$ by employing the $W_\mathrm{II}$ technique~\cite{Zaletel2015}. The finite temperature calculations are performed in a triangular cylinder of size $L=L_x\times L_y=15\times 3$.

To obtain the ground state of the system we employ the density matrix renormalization group (DMRG) algorithm. We perform simulations in a triangular cylinder of size $L=L_x\times L_y=15\times 6$ and we fix the maximum bond dimension of our MPS to $\chi=1024$. To capture the effects of a disordered on-site potential we have performed calculations in three different disorder realizations and average over them.

\section*{Acknowledgments}
The Authors acknowledge insightful discussions with Haydn Adlong, Bertrand Evrard and Liang Fu. This work was supported by the Swiss National Science Foundation (SNSF) under Grant Number 200021-204076. I.M. thanks support from Grant No. PID2020-114626GB-I00 from the MICIN/AEI/10.13039/501100011033 and Secretaria d'Universitats i Recerca del Departament d’Empresa
i Coneixement de la Generalitat de Catalunya, co-funded
by the European Union Regional Development Fund within
the ERDF Operational Program of Catalunya (project QuantumCat, Ref. 001-P-001644). E.D. acknowledges support from the ARO grant number W911NF-20-1-0163.
K.W. and T.T. acknowledge support from the JSPS KAKENHI (Grant Numbers 20H00354, 21H05233 and 23H02052) and World Premier International Research Center Initiative (WPI), MEXT, Japan.

\bibliography{references}

\end{document}